\documentclass[aip,rsi,amsmath,amssymb,reprint,floatfix]{revtex4-1}
\usepackage{graphicx}
\usepackage{float}
\usepackage{units}
\usepackage{dcolumn}
\usepackage{natbib} 

\begin{document}
\title{Optical readout of singlet fission biexcitons with photoluminescence detected magnetic resonance }
\author{Gajadhar Joshi}
\affiliation{National Renewable Energy Laboratory, Golden, Colorado 80401, United States}
\author{Ryan D. Dill}
\affiliation{Department of Chemistry, University of Colorado, Boulder, Colorado 80309, USA}
\author{Karl J. Thorley}
\affiliation{University of Kentucky Center for Applied Energy Research, Lexington, Kentucky 40511, USA}
\author{John E. Anthony}
\affiliation{University of Kentucky Center for Applied Energy Research, Lexington, Kentucky 40511, USA}
\affiliation{Department of Chemistry, University of Kentucky, Lexington, Kentucky 40506, USA}
\author{Obadiah G. Reid}
\affiliation{Renewable and Sustainable Energy Institute, University of Colorado, Boulder, Colorado 80309, USA}
\affiliation{National Renewable Energy Laboratory, Golden, Colorado 80401, United States}

\author{Justin C. Johnson}
\email{justin.johnson@nrel.gov}
\affiliation{National Renewable Energy Laboratory, Golden, Colorado 80401, United States}
\affiliation{Renewable and Sustainable Energy Institute, University of Colorado, Boulder, Colorado 80309, USA}
\date{\today}
\begin{abstract}
Molecular spin systems based on photoexcited triplet pairs formed via singlet fission (SF) are attractive as carriers of quantum information because of their potentially pure and controllable spin polarization, but developing systems that offer optical routes to readout as well as initialization is challenging. Herein, we characterize the electron spin magnetic resonance change in photoluminescence intensity for a tailored organic molecular crystal while sweeping a microwave drive up to \unit[10]{GHz} in a broadband loop structure. We observe resonant transitions for both triplet and quintet spin sublevel populations showing their optical sensitivity, and revealing zero-field parameters for each. We map the evolution of these spectra in both microwave frequency and magnetic field, producing a pattern of optically-detected magnetic resonance (ODMR) peaks. Fits to this data using a suitable model suggest significant spin polarization in this system with orientation selectivity. Unusual excitation intensity dependence is also observed, which inverts the sign of the ODMR signal for triplet features, but not for quintet. These observations demonstrate optical detection of the spin sublevel population dictated by SF and intermolecular geometry, and provide unique insight into the dynamics of triplet pairs.  
\end{abstract}

\maketitle
\section{\label{sec:level1} Introduction \protect\\}
Singlet fission (SF) is a photophysical process in which an excited singlet state (S$_1$) shares its energy with a ground state neighboring chromophore (S$_0$) to generate two triplet excited states (T$_1$+T$_1$) at distinct sites\cite{smith_singlet_2010}. The phenomenon of SF and its reverse process, triplet fusion (TF), were demonstrated in molecular solids by studying the magnetic field effects on photoluminescence (PL), and their theoretical account was well established on the basis of electron spin interactions\cite{singh_laser_1965,merrifield_fission_1969,merrifield_theory_1968}. Although the phenomenon became well known in the 1970s, it did not gain significant attention until the beginning of the 2000s. The resurgence of the field resulted from the recapitulated idea of exploiting singlet fission in photovoltaic devices to increase the efficiency of solar cells beyond conventional thermodynamic limits \cite{paci_singlet_2006,shockley_detailed_1961}. This notion exploits SF for exciton multiplication, i.e., absorption of a single photon produces two electron-hole pairs, which may subsequently evolve into free charge carriers\cite{hanna_solar_2006}. Recently, the observation of strongly correlated triplet pairs and the existence of quintet-state multiexcitons resulting from the SF process have opened new avenues for their application  in quantum information science (QIS)\cite{tayebjee_quintet_2017,bayliss_spin_2016,bayliss_probing_2020}. The correlated pairs are entangled spin states that in many cases possess well-defined initial polarization, which is one of the fundamental requirements for a practical qubit\cite{divincenzo_physical_2000}, and their coherent manipulation can be performed with microwave pulse sequences\cite{bayliss_spin_2016,petta_coherent_2005}.

Currently, several different approaches are under study for electron-spin based qubits, such as defect states in solids\cite{fuchs_gigahertz_2009,weber_2010,awschalom_quantum_2018,wolfowicz_quantum_2021}, molecular nanomagnets\cite{shiddiq_enhancing_2016,friedman_single-molecule_2010}, and quantum dots\cite{shulman_demonstration_2012}. One of the emerging systems for both quantum computation and sensing is the nitrogen vacancy (NV) center in diamond\cite{weber_quantum_2010,schirhagl_nitrogen-vacancy_2014}, due to its long spin coherence at room temperature and convenient optical initialization and readout. However, precise placement of defects at desired locations in crystal samples is extremely difficult, which limits scalability for potential quantum architectures. Organic molecular semiconductors are promising candidates to overcome these challenges using a 'bottom-up' approach. The fundamental properties of qubits are controlled at the molecular level during the synthesis of the desired compound, and long-range ordered assembly occurs through intermolecular interactions. In certain SF systems (vide infra), triplet population mimics the NV-center model in terms of ease of optical initialization and detection of the spin state, and is an excited state analogue to optically addressable ground-state spins of metal-ligand molecular compounds\cite{bayliss_optically_2020}. Of some concern is the potentially limited coherence time of molecular systems, given the large number of degrees of freedom present in the sample. However, there has already been significant progress toward improving the coherence times of correlated triplet pairs in molecular solids\cite{jacobberger_using_2022}, even at temperatures far above the ultracold conditions required for other systems, and the possibility of accessible ``clock" transitions has been theoretically predicted\cite{lewis_clock_2021}. 

Optically detected magnetic resonance (ODMR) is a naturally useful technique for judging emissive spin qubit behavior, as it relies on optical readout of the spin-polarized population upon microwave or RF stimulation. However, only a small subset of SF systems have triplet-pair states amenable to fluorescence detection - those with nearly degenerate S$_1$ and $^1$TT energy levels. Recent ODMR experiments on such an SF compound (TIPS-tetracene) in (poly)crystalline form have revealed meaningful information about the dynamics of triplet pairs\cite{yunusova_spin_2020,budden_singlet_2021,bayliss_geminate_2014}. However, the inability to observe the high-spin quintet species,\cite{budden_singlet_2021} and the complexity of spectra for crystals with several possible triplet-pair sites\cite{yunusova_spin_2020} have hindered a full development of the dynamical picture and evaluation of pure state production and readout. Therefore, other emissive SF systems are sought.  

The spin-Hamiltonian of two triplets residing at two molecular sites a and b is given by
\begin{multline*}
\hat{\mathcal{H}}= J\hat{\textbf{S}}_a.\hat{\textbf{S}}_b + \sum_{i=a,b} g \mu_B {\textbf{B}}_0.\hat{\textbf{S}}_i\\ + \sum_{i=a,b} [D({\hat{S}_{z,i}}^2-\frac{1}{3} S(S +1)) + E(\hat{S}_{x,i}^2 -\hat{S}_{y,i}^2)]
\label{ham} 
\end{multline*}

The first term represents the isotropic exchange coupling of triplet pair with exchange coupling strength $J$. The strength of the exchange interaction depends on the relative separation of center of triplet wave functions\cite{taffet_overlap-driven_2020}, and the sign of $J$ gives the type of interaction. Negative sign represents ferromagnetic coupling and positive sign represents antiferromagnetic coupling\cite{abraham_simple_2017,abraham_revealing_2021}. The magnitude of exchange coupling, $|J|$, can be determined from a magneto-photoluminescence experiment by observing the energy level crossing effect on photoluminescence (PL)\cite{bayliss_site-selective_2018}.  The second term describes the Zeeman interaction in an external field $B_\mathrm{0}$, and the third term describes the zero-field interaction within a triplet state in terms of axial and transverse zero-field parameters $D$ and $E$ respectively. 

In theoretical work based on a model triplet-pair spin Hamiltonian and nonadiabatic transition theory, known as the $JDE$ model\cite{smyser_singlet_2020}, it has been predicted that specific quintet sublevels can be populated selectively if the pair of molecules that undergoes SF fulfils certain conditions. One of the crucial criteria is the orientation of molecules in the dimer or crystal; the principal magnetic axis of the partners should be parallel with each other and with the direction of the external magnetic field. Motivated by this theoretical work, Rugg et.al., have applied  the $JDE$ model to a single crystal sample of a newly synthesized SF material 2-triethylsilyl-5,11-bis(triisopropylsilyl ethynyl) tetraceno[2,3-b]thiophene (TES TIPS-TT)\cite{Brandon_TES}, which possesses the requisite molecular alignment for pure spin polarization. Transient electron paramagnetic resonance (Tr-EPR) along with magneto-photoluminescence investigations of the single crystal sample have validated aspects of the $JDE$ model. Being an emissive material, TES TIPS-TT represents an opportunity to further refine models of spin polarization using ODMR methods, leading the way toward scalable production of optically addressable qubits. Fig.~\ref{fig1}a) shows the molecular structure of TES TIPS-TT with zero-field tensor axes in the inset. The zero-field tensors are assigned such that x-axis is along the major tetracene axis and the z-axis is perpendicular to the plane of the phenyl rings.

Here, we use broadband ODMR measurements of a polycrystalline powder sample of TES TIPS-TT at cryogenic temperatures to determine the optical sensitivity of various resonant transitions between spin sublevels of multiexcitonic states. The observed spectra are compared with the simulation results using EasySpin\cite{stoll_easyspin_2006} to assign the spin transitions responsible for particular features of the ODMR spectra. The ODMR spectra in these materials can be observed at zero external magnetic field, and their evolution with magnetic field in a broadband measurement gives more accurate information about fine-structure parameters and strength of interaction in the correlated pairs\cite{yunusova_spin_2020} than that of measurement at only one field. The laser intensity dependent ODMR spectra of triplets and quintets at zero-field suggest that there is an exciton density dependent interplay between the mechanisms of geminate vs nongeminate triplet-pair recombination\cite{bayliss_geminate_2014}.

\section{\label{sec:level1} Experimental Methods \protect\\}
\begin{figure}[t]
\centering
\includegraphics[scale=1]{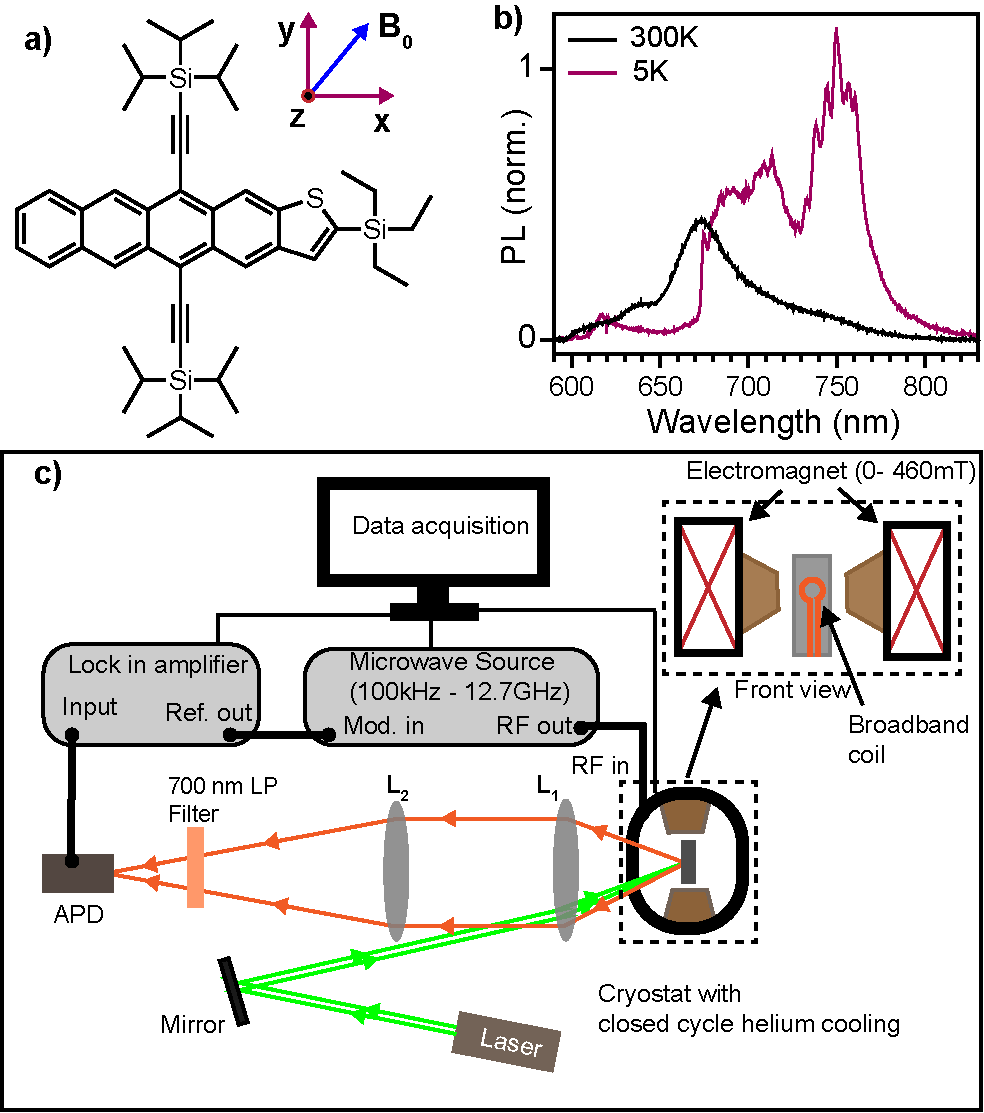}
\caption{Experimental scheme for the broadband ODMR study of triplet-pair states in a polycrystalline TES TIPS-TT powder. (a) Molecular structure of TES TIPS-TT and zero-field fine-structure tensor axes, with z-axis normal to the molecular plane of the conjugated ring system and applied magnetic field (\textbf{B}$_0$) (b) PL spectrum from sample at room-temperature and \unit[5]{K} when excited by \unit[532]{nm} diode laser. The spectra are collected with a \unit[600]{nm} long pass filter. (c) The experimental setup and data acquisition with lock-in detection scheme. The powder sample is placed inside the helium closed-cycle cryostat, and the optical assembly for photo-excitation and collection is outside the cryostat. Broadband RF excitation is achieved with a home-built non-resonant coil etched from a copper-coated FR-4 printed circuit board blank (see Supplementary Information).}
\label{fig1}
\end{figure}

A schematic of the broadband ODMR spectrometer is shown in Fig.~\ref{fig1}c). All measurements were performed at a temperature of \unit[5]{K} within a closed-cycle helium cryostat equipped with an electromagnet (Montana Instruments Cryostation with Magneto-Optic Module) with \unit[20]{mm} pole spacing. To achieve broadband RF excitation, we patterned a non-resonant copper loop of diameter about \unit[4]{mm}  on a printed circuit board (PCB). The TES TIPS-TT powder sample was sandwiched between two thin sapphire substrates to hold it in place close to the non-resonant loop. The sapphire substrates and the PCB board were pressed to the sample holder that is bolted on the cold finger of the cryostat for efficient cooling of the sample. Because of the limited space within the cryostat chamber, the optical excitation and collection optics are aligned externally. The output from a  \unit[532]{nm} diode laser (Thorlabs CPS532) is focused at the sample with a biconvex lens L$_1$ and the PL of the sample is also collimated by the same lens. A \unit[50]{mm} f/0.95 C-mount camera lens L$_2$ (Thorlabs MVL50HS) focuses the PL onto an avalanche photodetector (APD, Thorlabs APD430A), which converts optical signal into electrical signal. A \unit[700]{nm} long pass filter is placed in front of the APD to reduce collection of scattering from the incident beam. The output of the APD is routed to the input channel of a lock-in amplifier (Zurich Instruments UHFLI; \unit[600]{MHz} bandwidth) and continuous wave (cw) ODMR  data are acquired by amplitude modulation of the microwave source (Rohde and Schwarz SMB100A) at \unit[200]{Hz}. For the excitation density measurements, a collimating lens assembly (not shown in Fig.~\ref{fig1}c)) is inserted in the path between the laser and mirror to control the excitation spot size.  With the custom pole tips, the experiment is able to provide a magnetic field up to \unit[460]{mT}. The magnetic field at the sample position is calibrated with an ODMR measurement of a thin film of the pi-conjugated poly[2-methoxy-5-(2'-ethylhexyloxy)-1,4-phenylene vinylene] (MEH-PPV)\cite{baker_robust_2012}. For zero-field measurements, the true zero-field at the sample position is confirmed by observing zero-field ODMR of negatively charged NV-center crystal sample at \unit[2.87]{GHz}\cite{sasaki_broadband_2016}. The data acquisition and magnet control are accomplished with custom Python code developed in-house. 

\section{\label{sec:level1} Results \protect\\}
PL spectra of the TES TIPS-TT powder at room temperature and \unit[5]{K} are shown in Fig.~\ref{fig1}b). Cooling the sample causes the emergence of a strong emission band with sharp features centered around \unit[750]{nm}. The spectra shown in Fig.~\ref{fig1}b) were collected with a \unit[600]{nm} long-pass filter to block the laser scatter; for all other ODMR measurements, a \unit[700]{nm} long-pass filter was used to isolate the red-shifted emission. Similar enhancement of red-shifted PL in other SF systems has been assigned to $^1$TT emission\cite{bossanyi_emissive_2021}. The Herzberg-Teller mechanism could allow such emission, and is enhanced in TES TIPS-TT by the similarity 
between the S$_1$ and $^1$TT energies. The overall PL decreases as magnetic field is increased, followed by a slow rise towards a plateau at our experimental field limit (see supplementary figure S1). This behavior is commonly observed in SF materials at relatively low magnetic field range\cite{bayliss_site-selective_2018,yago_magnetic_2016}. 

Due to the exchange interaction, $J$, that couples the triplets in a pair, the multiexcitonic states, quintet ($S$=2) and triplet ($S$=1), are separated from the singlet state ($S$=0) by 3$|J|$ and $|J|$, respectively \cite{bayliss_site-selective_2018,tayebjee_quintet_2017}. The anisotropic exchange coupling between the triplets is typically small in comparison to the isotropic exchange coupling term, thus is ignored in the spin Hamiltonian of coupled pairs. The anisotropic term, however, does cause a slight deviation of the quintet zero-field parameter, $D_\mathrm{Q}$, from the theoretical limit\cite{bayliss_geminate_2014}, $D_\mathrm{T}$/3.  Quintet and triplet states are further split by intra-triplet zero-field splittings. The magnetic dipolar interaction between the triplets of the correlated triplet pairs produced by SF is typically on the order of a few MHz\cite{wang_magnetic_2015}, much smaller than the isotropic coupling term, and is hence neglected in the spin Hamiltonian discussed above\cite{tayebjee_quintet_2017}. 
\begin{figure}[b]
\centering
\includegraphics[scale=1]{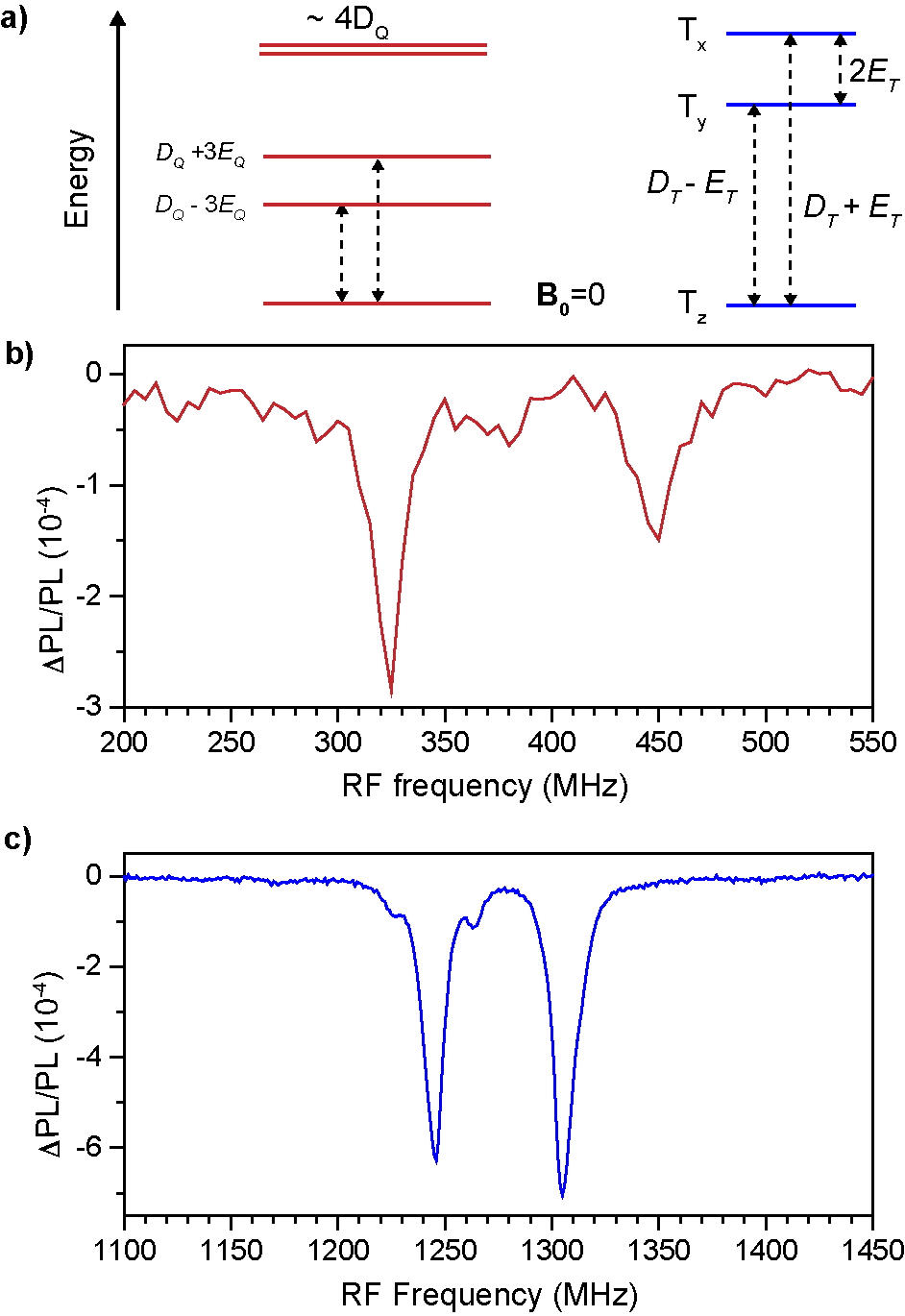}
\caption{Zero-field ODMR of quintets and triplets. (a) The energy splitting of quintet and triplet spins at zero external field. The dotted arrows show the observed transitions between spin sublevels.(b) Zero-field ODMR of quintet states provides the quintet fine-structure parameters $|D_\mathrm{Q}|$ and $|E_\mathrm{Q}|$.(c) Triplet ODMR  gives the zero-field parameters $|D_\mathrm{T}|$ and $|E_\mathrm{T}|$. The transition between T$_\mathrm{y}$ and T$_\mathrm{x}$ is not shown here. The weaker transitions flanked on both sides of the triplet peak at \unit[1244]{MHz} have not been assigned.}
\label{fig2}
\end{figure}
Fig.~\ref{fig2}a) shows the zero-field splittings of quintet and triplet states, revealing unique transitions within the different spin manifolds. Fig.~\ref{fig2}b,c) shows quintet and triplet zero-field ODMR spectra, respectively. Previously observed ODMR at zero-field has shown the lowest ODMR transition for quintets and triplets occur at frequencies $D_\mathrm{Q}\pm 3E_\mathrm{Q}$ and $D_\mathrm{T} \pm E_\mathrm{T}$ respectively\cite{yunusova_spin_2020}.  The spin fine-structure parameters of quintets and triplets for TES TIPS-TT are determined to be $|D_\mathrm{Q}|$= \unit[386]{MHz}, $|E_\mathrm{Q}|$= \unit[20]{MHz} and $|D_\mathrm{T}|$= \unit[1273]{MHz}, $|E_\mathrm{T}|$= \unit[30]{MHz} respectively. The sign of these fine-structure parameters cannot be determined from ODMR alone, and hereafter we adopt positive $D_\mathrm{Q}$ and $D_\mathrm{T}$, and negative $E_\mathrm{Q}$ and $E_\mathrm{T}$, as determined from Tr-EPR and EPR methods\cite{Brandon_TES,yarmus_epr_1972}. We have observed strong transitions between T$_\mathrm{y}$  and T$_\mathrm{x}$ (see supplementary figure S2), which indicates a large steady-state spin polarization of triplet states in the zero-field basis of the molecules.  We have not observed the quintet transition between $|D_\mathrm{Q}|$ - 3$|E_\mathrm{Q}|$ and $|D_\mathrm{Q}|$ + 3$|E_\mathrm{Q}|$ at frequency 6$E_\mathrm{Q}$. 

\begin{figure}[t]
\centering
\includegraphics[scale=1]{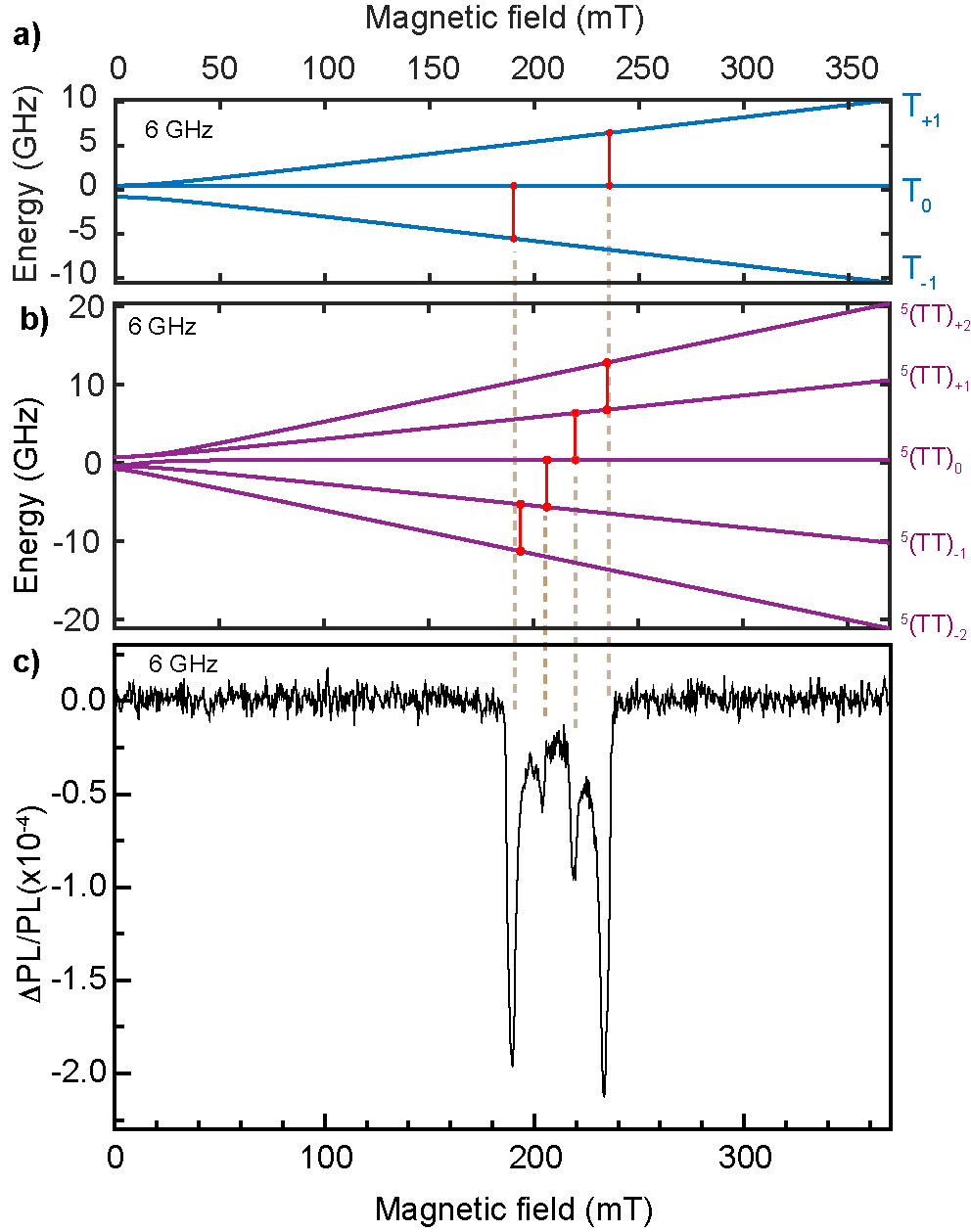}
\caption{Field swept ODMR of TES TIPS-TT at \unit[6]{GHz} and comparison of spectral positions in EasySpin simulation. (a) Energy levels of uncoupled triplets as a function of external magnetic field, and calculated EPR transitions at \unit[6]{GHz} with the zero-field parameters $D_\mathrm{T} = \unit[1273]{MHz}$ and $E_\mathrm{T}= -\unit[30]{MHz}$ extracted from the experiment.(b) Quintet energy levels of coupled triplets as a function of applied magnetic field for $D_\mathrm{Q} = \unit[386]{MHz}$ and $E_\mathrm{Q}= -\unit[20]{MHz}$. The possible EPR transitions at \unit[6]{GHz} are marked by red vertical lines. (c) The experimental ODMR data at 6GHz showing only four transitions. The inner transitions are from the quintet state transition $^{5}$(TT)$_{-1}$ to  $^{5}$(TT)$_0$ and $^{5}$(TT)$_{+1}$ to $^{5}$(TT)$_0$ while the outer stronger peaks are assigned to triplet transitions  T$_{-1}$ to T$_0$ and T$_{+1}$ to T$_0$}
\label{fig3}
\end{figure}
\begin{figure*}[t]
    \centering
    \includegraphics[width=\textwidth]{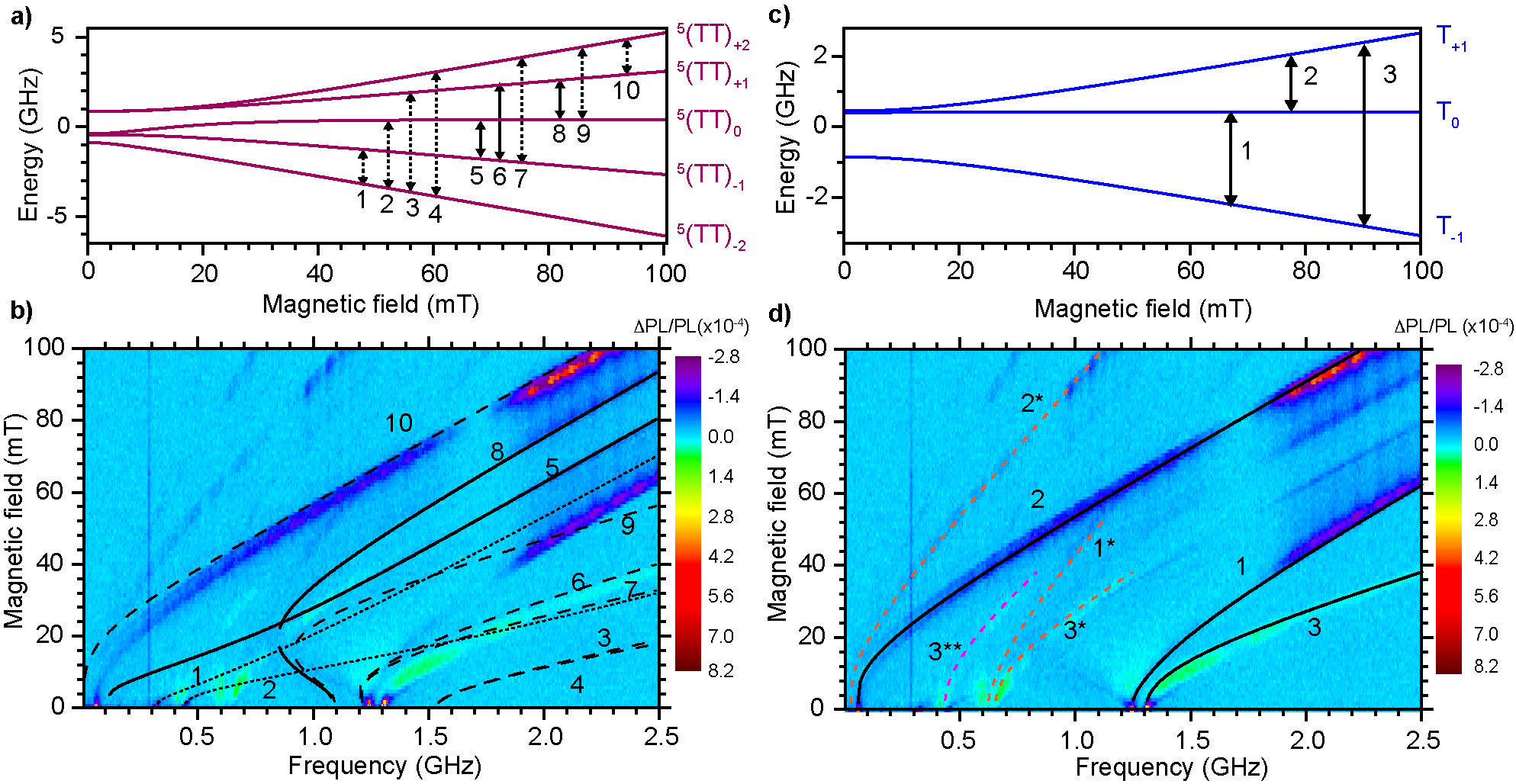}
    \caption{Broadband ODMR at low magnetic field range. (a) Energy levels of quintet states as a function of magnetic field. There are 10 possible transitions among various levels, labeled on the plot. The dashed transitions are not observed in our broadband ODMR measurement. (b) The overlay of broadband ODMR measurement with the calculated EPR transition frequencies in EasySpin for quintet state (S=2) with experimentally obtained zero-field parameters $D_\mathrm{Q}$ and $E_\mathrm{Q}$, and magnetic field orientation of [45$^{\circ}$, 90$^{\circ}$, 0$^{\circ}$] with respect to the molecular frame. A narrow vertical line near \unit[0.3]{GHz} is an artifact in the measurement. The solid curves correlate with observed transitions in the measurement. The dashed curves are not observed in the ODMR measurements. The dotted curves only appears to match at zero field but does not appear at higher magnetic field. (c) Triplet energy levels as a function of magnetic field. The possible transitions are marked from 1-3. (d) The transition frequencies for triplet levels are calculated and overlaid in the broadband measurement with the same orientation as quintets.  Several apparent lines in the spectra that are not accounted for in the simulations appear at integer multiples of identified transitions. These are attributed to harmonic artifacts in the microwave drive and are represented by dashed lines. The second and third harmonics of corresponding transitions are labeled with `*' and `**' respectively.}
    \label{fig4}
\end{figure*}
\begin{figure}[t]
\centering
\includegraphics[scale=1]{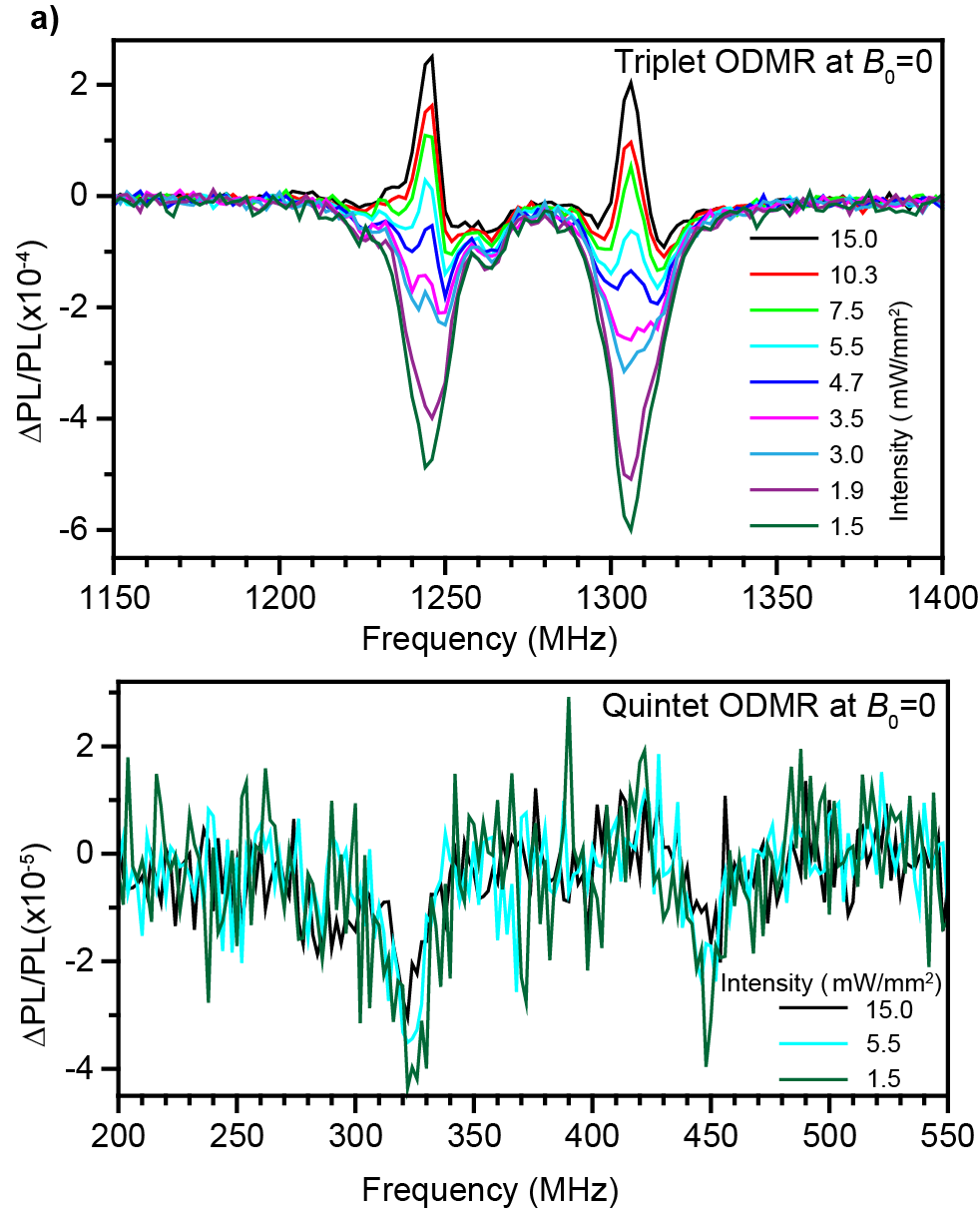}
\caption{Laser intensity dependent triplet and quintet ODMR spectra at zero external field. (a) Triplet and (b) quintet ODMR at zero-field for various laser intensities.}
\label{fig5}
\end{figure}

The zero-field parameters guide the energy levels of different spin-sublevels in an external magnetic field, and hence quintet and triplet sublevels can be further distinguished at higher magnetic fields i.e., $B_0\gg D_\mathrm{T}/g\mu_B$. A field-swept ODMR spectrum at \unit[6]{GHz} RF drive frequency, Figure 3c, appears to be a superposition of $^5$TT and T$_1$ transitions. These can be distinguished and assigned properly by comparing the results with spectral simulations\cite{stoll_easyspin_2006}. Due to the parallel intermolecular arrangement of molecules in TES TIPS-TT crystals, the transitions from coupled triplet states $^3$TT are not expected to be directly populated from $^1$TT states \cite{smyser_singlet_2020}. Their strong overlap with T$_1$ transitions also makes them difficult to distinguish, and we likely only observe the transitions from coupled quintet states $^5$TT and free triplet states T$_1$. When $|J| \gg D_\mathrm{T}$, the triplet and quintet manifolds are well separated and hence their evolution and transition frequencies can be determined independently. Therefore, we simulated EPR transitions for quintet with $S$ = 2, and zero-field parameters $D_\mathrm{Q}$ and $E_\mathrm{Q}$ as determined above. The triplet EPR transitions are calculated with $S$=1, and triplet zero-field parameters $D_\mathrm{T}$ and $E_\mathrm{T}$. Fig.~\ref{fig3}c) shows the observed ODMR signal at \unit[6]{GHz}, which is comprised of two outer stronger transitions and two relatively weaker inner transitions. To understand the origin of these transitions, we simulated an energy level plot with EPR transitions in EasySpin at \unit[6]{GHz} using B$_0$ parallel to the x-y molecular plane at a 45$^\circ$ angle relative to the x-axis (see EasySpin simulation section in SI).  Fig.~\ref{fig3}a) is the energy level plot of the uncoupled triplet state, and the corresponding EPR transitions are marked by red vertical lines. The simulation is performed with above determined zero-field parameters $D_\mathrm{T}$, $E_\mathrm{T}$. Fig.~\ref{fig3}b) is the energy level plot for the quintet state at \unit[6]{GHz} obtained with quintet zero-field parameters, $D_\mathrm{Q}$, $E_\mathrm{Q}$. EPR and ODMR spectra both result from spin transitions between two allowed spin sublevels, but they have different sensitivity because not all transitions cause a change in PL intensity of the samples. Nonetheless, \unit[6]{GHz} ODMR clearly shows the inner, weaker, features are from the quintet state ($^5$TT) transitions between $m_s =\pm 1$ and $m_s=0$, and the strong outer transitions are from the uncoupled triplet state (T$_1$), between $m_s =\pm 1$ and $m_s=0$. The strong relative amplitude of the free triplet associated transitions could be caused by several factors, but the primary factor is probably the much larger time-integrated population and spin-lattice relaxation lifetime of free triplets compared to quintets\cite{Brandon_TES}. 
Although spin sublevels are well separated at higher magnetic fields,  their evolution at intermediate fields (B$_0 \sim D_\mathrm{T}/g\mu_B$) is not straightforward. To understand the evolution of spin sublevels in correlated pairs, we carried out frequency sweep measurements with increasing magnetic fields up to \unit[100]{mT}. The observed magnetic resonance patterns are compared to simulation of resonant EPR transitions\cite{stoll_easyspin_2006} for coupled and uncoupled triplets. Fig.~\ref{fig4}b) represents the experimental data with overlay of possible EPR transitions for the quintet state with zero-field parameters $D_\mathrm{Q}$ and $E_\mathrm{Q}$, obtained experimentally. To determine the transitions that are observed in ODMR measurements, the energy level diagram of quintet states is plotted in Fig.~\ref{fig4}a). In total there are ten possible EPR transitions. The transitions that are seen in the measurement are plotted with dark lines, and transitions that are not observed are plotted in dashed lines. The two dotted curves in Fig.~\ref{fig4}b) are observed near zero-field. These transitions disappear quickly with increasing magnetic field, and do not appear at high field. It is evident that the inner transitions in the ODMR originate from the quintet spin state from the $m_s =0$ to the $m_s =\pm 1$ levels. The stronger transitions in the ODMR cannot be explained by quintet states, and thus we compare the observed spectra with EPR transitions of uncoupled triplet states. Fig.~\ref{fig4}c,d) shows the possible transitions and the overlay of calculated transition frequencies on top of experimental data. 
The ODMR spectrum of the strong triplet transitions at zero-field depends strongly on the intensity of the laser excitation source, Fig.~\ref{fig5}a). At low excitation densities, around \unit[1.5]{mW/mm$^2$}, the ODMR signals indicate quenching of steady state PL ($\Delta$PL/PL $< 0$) when RF is present. But as the excitation density increases to about \unit[15]{mW/mm$^2$}, the signal changes sign, indicating \textit{enhancement} of PL. This behavior is unprecedented in the literature for triplet ODMR at zero field in acenes. We did not observe such behavior for quintet ODMR at zero-field as shown in Fig.~\ref{fig5}b).

\section{\label{sec:level1} Discussion \protect\\}

The apparent orientation selectivity for \textbf{B$_0$} parallel to the molecular x-y plane, necessary to match predicted to observed transitions, is \textit{a priori} unexpected for powder-like TES TIPS-TT. Prior work on a disordered polycrystalline TIPS-tetracene film suggested that \textbf{B$_0$}$\parallel \hat{\mathrm{z}}$ is preferred in ODMR, based on the notion that free triplets experience the slowest spin-lattice relaxation under this condition, thus microwave pumping is much more effective than at \textbf{B$_0$}$\bot\hat{\mathrm{z}}$. However, our prior work suggests minima in spin-relaxation rates for both \textbf{B$_0$} parallel and perpendicular to $\hat{\mathrm{z}}$ for TES TIPS-TT, with faster rates for orientations in between\cite{Brandon_TES}. \textbf{B$_0$}$\bot \hat{\mathrm{z}}$ has the slowest rate and therefore may contribute the largest signal in time-integrated ODMR. The physical origin of the orientation selectivity is not clear but may be related to the contribution to spin-lattice relaxation from exciton diffusion. Based on its crystal structure, TES TIPS-TT is likely to have fast and highly anisotropic diffusion of triplets, unlike TIPS-tetracene, which does not have fully aligned molecules in the unit cell and therefore shows slow triplet transport, especially at low temperature\cite{stern_vibronically_2017}.

\begin{figure}[t]
\centering
\includegraphics[scale=1]{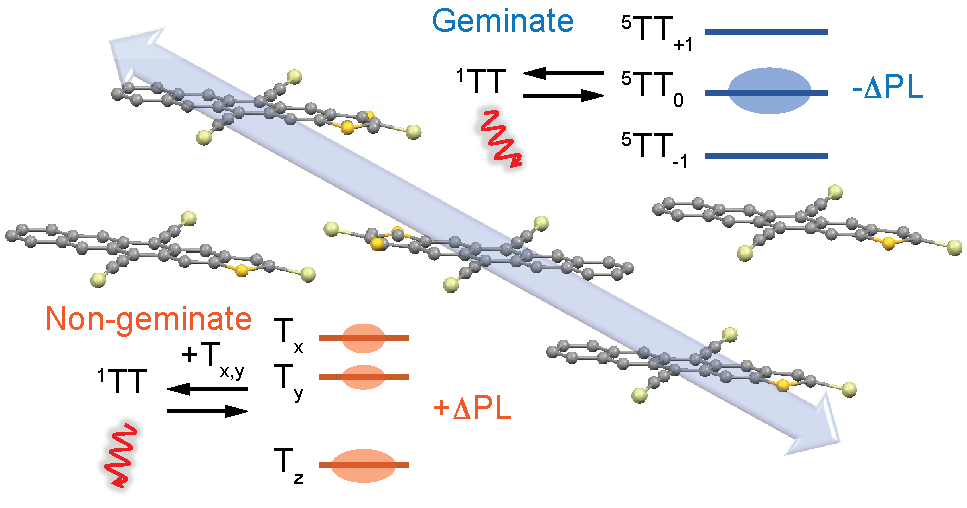}
\caption{Schematic depiction of the mechanism of geminate and nongeminate recombination in TES TIPS-TT based on the known crystal structure. The fast transport axis is shown (blue double arrow). Spin polarization in the inset diagrams is suggested to be due to SF selection rules (geminate case) and Boltzmann distribution (nongeminate case), modified by the depletion of singlet character states through fast transitions to the emissive $^1$TT, facilitated by T+T encounters.  }
\label{fig6}
\end{figure}
It has been previously established that the ODMR signals in singlet fission materials are comprised of both geminate and nongeminate triplet-triplet interactions\cite{bayliss_geminate_2014,budden_singlet_2021}. The triplet pair generated at a molecular site in the SF process is born as $^1$TT, and the constituent triplets can subsequently recombine without interaction of separately generated triplets. Such a process is geminate triplet pair recombination and not expected to be intensity dependent. In the non-geminate triplet-triplet recombination process, triplets born through spatially separate SF events can encounter each other through hopping in the material, resulting in recombination and delayed emission through the singlet excited state channels. The sign of zero-field ODMR has been suggested to be different for these two mechanisms\cite{bayliss_geminate_2014} due to the degree of singlet character in the participating species for each process as depicted in Fig.~\ref{fig6}. Geminate triplet pairs retain singlet memory prior to re-encounter and thus lose emissivity upon stimulation away from these states. The opposite is true for nongeminate triplet-triplet encounters, where the population distribution typically becomes Boltzmann-like over time, funneling population toward the ``dark'' energy levels, particularly at low temperature. Thus, resonant microwave driving would restore more singlet character and higher emissivity. This remains plausible but unproven in our case because we did not perform a rigorous temperature dependence to alter Boltzmann populations. We suggest an alternative mechanism in which enhanced recombination of the singlet character population depletes these states ( Fig.~\ref{fig6}), and thus resonant transitions within the darkened population restore additional ``bright'' states. We consider T$_1$+T$_1$ encounters in TES TIPS-TT to be largely geminate at low excitation intensity, as separation and recombination likely occur within a one-dimensional stack of aligned molecules as shown in Fig.~\ref{fig6}, retaining the high probability of spin singlet memory. For quintets that remain in the strong exchange coupling regime, the interaction is naturally geminate.  Both of these mechanisms lead to negative ODMR. 

  However, as intensity is increased, the triplet signals change from negative to positive ODMR, suggesting a change to a nongeminate process.  As triplet-triplet recombination in this case is naturally bimolecular, its rate will increase with higher light intensity.  As can be seen in Fig.~\ref{fig5}a), the sharp positive feature emerges around \unit[3]{mW/mm$^2$}, before becoming dominant and inverting the overall ODMR sign by about \unit[6]{mW/mm$^2$}. As triplet density increases with excitation intensity, nongeminate T$_1$-T$_1$ interactions involving re-encounters are enhanced (Fig.~\ref{fig6}), eventually outweighing the geminate process and effectively cancelling the negative ODMR signature. The intensity dependence of the quintet ODMR signals is linear and does not change in sign.  This is consistent with quintet signals arising predominately from geminate pairs. The narrower lineshape of the features at high intensity may also suggest they are associated with select recombination sites preferred for bimolecular triplet-triplet encounters. In this sense the full distribution of triplet sites becomes dominated by a smaller portion of available nearest-neighbor geometries, and the inhomogeneously broadened linewidth is reduced.

We note that some predicted quintet peaks lose intensity sharply as magnetic field increases from zero-field (dotted lines in Figure 4b). Upon low to intermediate field application, the zero-field triplet-pair states possess significant singlet character across the entire manifold. This may reduce the efficacy of RF-stimulated transitions in terms of distributing population between bright and dark states and quench ODMR. As the high-field regime is approached within the quintet manifold (\textbf{B}$_0$ $\gg D_\mathrm{T}$), only $^5$(TT)$_0$ maintains a direct pathway to $^1$TT. Additionally, inversion of the ODMR sign of triplets with higher excitation intensity is not observed at high magnetic field. We postulate that field enforces a strong T$_0$ spin polarization (not strictly Boltzmann) that renders the geminate vs. nongeminate mechanism similar in terms of the involvement of triplet sublevels. Therefore, microwave-stimulated transitions away from T$_0$ reduce the probability of a fast pathway to the partially emissive $^1$TT equally at all intensities. We note that unlike oriented single crystal studies\cite{Brandon_TES}, powders do not result in macroscopic alignment of all molecules at the same orientation with respect to the field (e.g., \textbf{B}$_0$ parallel to z-molecular axis), thus allowing some $m_s=\pm 2$ quintet population; however, single quantum transitions from these sublevels are unlikely to influence emission because both participant levels are dark. Thus, several theoretically possible transitions are not observed at high field (Figure 4b, dashed).

\section{\label{sec:level1} Conclusion \protect\\}

We have shown that both individual triplets and triplet pairs with well-defined spin sublevel populations can be detected through the strong emission of the $^1$TT state in a molecular semiconductor designed both energetically and structurally to exhibit simple and strong magnetic resonance transitions. The ODMR signal is consistent with spin polarization dictated by SF and the involvement of transitions to and from the bright $^1$TT state . The unexpected orientation selectivity and power dependence can be explained by the unique crystal structure of TES TIPS-TT, which contains parallel stacks of aligned molecules. This demonstrates generation and optical readout of highly spin-polarized T$_0$ and $^5$(TT)$_0$ states in a synthetic system that was designed for this purpose, highlighting the potential of such systems as tailored bottom-up alternatives to nitrogen vacancy centers in diamond and other defect states generated by stochastic processes.  Future work on single crystals and derivatives that further isolate dimer-like interactions may slow spin relaxation and result in high fidelity gate operations driven by microwaves and detected with light.

\section{\label{sec:level1} Supplemental Material \protect\\}

See supplementary material for further magneto-photoluminescence data, field-swept ODMR spectra, and additional information about ODMR simulations.

\section{\label{sec:level1} Acknowledgments \protect\\}

This work was authored by Alliance for Sustainable Energy, LLC, the manager and operator of the National Renewable Energy Laboratory for the U.S. Department of Energy (DOE) under Contract No. DE-AC36-08GO28308. Funding provided by U.S. Department of Energy, Office of Basic Energy Sciences, Division of Chemical Sciences, Biosciences, and Geosciences. 
The views expressed in the article do not necessarily represent the views of the DOE or the U.S. Government. The U.S. Government retains and the publisher, by accepting the article for publication, acknowledges that the U.S. Government retains a nonexclusive, paid-up, irrevocable, worldwide license to publish or reproduce the published form of this work, or allow others to do so, for U.S. Government purposes. 

\section{\label{sec:level1} References \protect\\}
\bibliography{references}

\end{document}